\documentclass[10pt]{article}
\usepackage{amsmath}
\usepackage{latexsym}
\usepackage[normalem]{ulem}
\usepackage{soul}
\usepackage{array}
\usepackage{extarrows}
\usepackage{enumitem}
\usepackage{graphicx}

\usepackage{subfig}
\usepackage{wrapfig}
\usepackage{txfonts}
\usepackage{wasysym}
\usepackage{enumitem}
\usepackage{adjustbox}
\usepackage{ragged2e}
\usepackage[svgnames,table]{xcolor}
\usepackage{tikz}
\usepackage{longtable}
\usepackage{changepage}
\usepackage{setspace}
\usepackage{hhline}
\usepackage{multicol}
\usepackage{tabto}
\usepackage{float}
\usepackage{multirow}
\usepackage{makecell}
\usepackage{fancyhdr}
\usepackage[toc,page]{appendix}
\usepackage[hidelinks]{hyperref}
\usetikzlibrary{shapes.symbols,shapes.geometric,shadows,arrows.meta}
\tikzset{>={Latex[width=1.5mm,length=2mm]}}
\usepackage{flowchart}
\usepackage[paperheight=11.0in,paperwidth=8.5in,left=0.95in,right=0.95in,top=0.95in, textheight = 9.2in,headheight=0.75in,headsep=\baselineskip]{geometry}

\usepackage[utf8]{inputenc}
\usepackage[T1]{fontenc}
\usepackage{tikz}
\usetikzlibrary{arrows}
\usepackage{verbatim}
\usetikzlibrary{shapes, arrows,positioning}
\usepackage{tikz}
\usetikzlibrary{shapes,arrows.meta,decorations}
\TabPositions{0.5in,1.0in,1.5in,2.0in,2.5in,3.0in,3.5in,4.0in,4.5in,5.0in,5.5in,6.0in,}
\usepackage[labelfont=bf]{caption}
\usepackage{titlesec}
\usepackage{cite}

\titleformat{\section}
  {\normalfont\fontsize{12}{15}\bfseries}{\thesection}{1em}{}
\titleformat{\subsection}
  {\normalfont\fontsize{10}{15}\bfseries}{\thesubsection}{0.5em}{}

\renewcommand{\thesection}{\arabic{section}.}
\renewcommand{\thesubsection}{\arabic{section}.\arabic{subsection}.{}}

\titlespacing\section{0pt}{12pt plus 4pt minus 2pt}{0pt plus 2pt minus 2pt}
\titlespacing\subsection{0pt}{12pt plus 4pt minus 2pt}{0pt plus 2pt minus 2pt}
\titlespacing\subsubsection{0pt}{12pt plus 4pt minus 2pt}{0pt plus 2pt minus 2pt}
\setlistdepth{9}
\renewlist{enumerate}{enumerate}{9}
		\setlist[enumerate,1]{label=\arabic*)}
		\setlist[enumerate,2]{label=\alph*)}
		\setlist[enumerate,3]{label=(\roman*)}
		\setlist[enumerate,4]{label=(\arabic*)}
		\setlist[enumerate,5]{label=(\Alph*)}
		\setlist[enumerate,6]{label=(\Roman*)}
		\setlist[enumerate,7]{label=\arabic*}
		\setlist[enumerate,8]{label=\alph*}
		\setlist[enumerate,9]{label=\roman*}

\renewlist{itemize}{itemize}{9}
		\setlist[itemize]{label=$\cdot$}
		\setlist[itemize,1]{label=\textbullet}
		\setlist[itemize,2]{label=$\circ$}
		\setlist[itemize,3]{label=$\ast$}
		\setlist[itemize,4]{label=$\dagger$}
		\setlist[itemize,5]{label=$\triangleright$}
		\setlist[itemize,6]{label=$\bigstar$}
		\setlist[itemize,7]{label=$\blacklozenge$}
		\setlist[itemize,8]{label=$\prime$}

\usepackage{nomencl}
\usepackage{etoolbox}
\makenomenclature
\renewcommand\nomgroup[1]{%
  \item[\bfseries
  \ifstrequal{#1}{I}{Sets and Indices}{%
  \ifstrequal{#1}{P}{Parameters}{%
  \ifstrequal{#1}{V}{Variables}{}}}%
]}

\usepackage[parfill]{parskip}

\begin{document}

\chead{}
\lhead{
\textit{}
}

\begin{Center}
{\fontsize{16pt}{19.2pt}\selectfont \textbf{Impacts of Privately Owned Electric Vehicles on Distribution System Resilience: A Multi-agent Optimization Approach}\\}
\end{Center}\par

\begin{Center}
{Sina Baghali, Zhaomiao Guo}\\
{Department of Civil, Environmental, and Construction Engineering, University of Central Florida}
\end{Center}\par






\begin{Center}
{\fontsize{12pt}{19.2pt}\selectfont \textbf{Abstract}}
\end{Center}

We investigate the effects of private electric vehicles (EVs) on the resilience of distribution systems after disruptions. We propose a framework of network-based multi-agent optimization problems with equilibrium constraints (N-MOPEC) to consider the decentralized decision making of stakeholders in transportation and energy systems. To solve the high-dimensional non-convex problem, we develop an efficient computational algorithm based on exact convex reformulation. Numerical studies are conducted to illustrate the effectiveness of our modeling and computational approach and to draw policy insights. The proposed modeling and computational strategies could provide a solid foundation for the future study of power system resilience with private EVs in coupled transportation and power networks.

\textbf{Keywords}\\
Distribution system resilience, electric vehicles, incentive design, multi-agent optimization, convex reformulation.
\section{Introduction}
Power system resilience reflects the versatility of a system to withstand and recover rapidly from unexpected disruptions \cite{khazeiynasab2020resilience,lei2018routing}. An increasing market penetration of private electric vehicles (EVs) provides new opportunities for enhancing power system resilience due to their mobility and fast regulating characteristics\cite{liu2016ev}.  In this paper, we will investigate the possible impacts of private EVs on assisting the restoration process of distribution systems (DSs).

Current research on DS restoration with EVs have focused on a centralized perspective without considering the non-cooperative travel and charging behavior of private EVs, as well as other decentralized power system stakeholders \cite{haggi2019review}. For example, researchers in \cite{lei2018routing,xu2019resilience} consider EVs as a type of mobile energy sources (MESs) along with mobile energy storage systems and mobile generators in DS restoration. Authors in \cite{jamborsalamati2019enhancing} investigate the centralized real time management of DS restoration with EV aggregators and distributed generators (DGs). In all the mentioned studies, a centralized entity, i.e., distribution system operator (DSO) is in full control of DS and restoration strategies, making decisions for DGs and EVs. However, centralized controlling techniques are proven to be challenging in terms of costly communication infrastructure and single-point failures \cite{nejad2019distributed}. In addition, privately-owned EVs, as well as other energy providers in DSs, may have their own objectives and can not be considered as manageable elements.

Moreover, due to large-scale private EVs participating in Vehicle-to-Grid (V2G) services, power and transportation systems become more interdependent. However, most of the existing studies on power system restoration with EVs take a power-system-centric perspective and completely neglect transportation systems \cite{jamborsalamati2019enhancing,momen2020using,sun2018optimal} or simplify the modeling of transportation systems by considering constant predefined travel time on routes \cite{lei2018routing,yao2018transportable,yao2019resilient}. This stream of literature overlooks the inherent relationship between traffic distribution and travel time that are critical especially when large-scale EVs are available. As a consequence, spatial and temporal interdependence between transportation and power systems can not be properly investigated.

We address the aforementioned gaps by proposing a a framework of network-based multi-agent optimization problems with equilibrium constraints (N-MOPEC) in a coupled distribution and transportation system. The main contribution is two-fold: (1) The modeling framework captures the decentralized behavior of stakeholders during distribution system restoration process and the spatial and temporal interdependence between transportation and power systems, which allows for rigorous system analyses and optimal V2G incentives design. (2) To facilitate large-scale computation, we reformulate the multi-agent optimization problems as an exact convex optimization problem, which can be efficiently solved by commercial nonlinear solvers. 
\vspace{-0.2cm}
\section{Mathematical Modeling}\label{sec:MAth_modeling}
We consider the following stakeholders: DSO, EV drivers, charging station aggregator (CSA), and DG owners for distribution system resilience. The decentralized decision making of each stakeholder is modeled in this section. Throughout this paper, set $\mathcal{I}$ represents the DS nodes, and $\mathcal{I}^{\mathrm{CS}}$, $\mathcal{I}^\mathrm{DG}$, and $\mathcal{I}^L$ are subsets of this set representing charging station nodes, DG nodes, and load nodes respectively. The time period of 24 hours are included in set $\mathcal{T}$, and $\mathcal{L}$ represents the set of distribution lines. Transportation networks are presented as a directed graph $\mathcal{G(\mathcal{N}, \mathcal{A})}$, within which EV drivers starting from set $\mathcal{R}$ select from charging station set $\mathcal{S}$ to provide grid restoration services and/or to charge.

\underline{{\bf DG Owners Modeling: }}Each DG $i$ ($\in \mathcal{I}^{DG}$) determines its generation quantity $p_{i,t}^{DG}$ for each time step $t$ ($\in \mathcal{T}$) to optimize its profits. Because individual DG does not have market power to influence the locational electricity prices $\rho_{i,t}$, the decision making of all DG owners can be aggregated into a single optimization problem, as formulated in model (\ref{mod:sp}).
    \begin{subequations}\label{mod:sp}
    \small{
		\begin{align}
		\max_{\substack{\boldsymbol{p}^{DG}} \geq \boldsymbol{0}} &  \sum_{i \in \mathcal{I}^{DG}}\sum_{t \in \mathcal{T}} \big( \rho_{i,t} p_{i,t}^{DG} - C_i(p_{i,t}^{DG}) \big) 
		\label{obj:DG} \\
		\text{ \ s.t.} & \ \underbar{P}_{i,t}^{DG} \leq p_{i,t}^{DG} \leq \bar{P}_{i,t}^{DG}, \ \forall i \in \mathcal{I}^{DG}, t \in \mathcal{T}  \label{cons:DG_max}
		\end{align}}
	\end{subequations}
	
\setlength{\headsep}{0in}
\fancyhf{}
\chead{ \begin{Center}
\textit{S. Baghali and Z. Guo}
\end{Center}}

Objective (\ref{obj:DG}) maximizes the profits of DG owners calculated as the total revenue $\sum_{i \in \mathcal{I}^{\mathrm{DG}}}\sum_{t \in \mathcal{T}} \rho_{i,t} p_{i,t}^{DG}$ subtracting the total production costs $\sum_{i \in \mathcal{I}^{\mathrm{DG}}}\sum_{t \in \mathcal{T}} C_i(p_{i,t}^{DG})$. We assume $C_i(\cdot)$ to be a convex function with respect to $p_{i,t}^{DG}$ \cite{yan2018robust}. Constraint (\ref{cons:DG_max}) determines the upper and lower bounds ($\bar{P}_{i,t}^{DG}$/$\underbar{P}_{i,t}^{DG}$) for power generation at each DG node $i$ for time $t$. When DGs can be disconnected from the systems, $\underbar{P}_{i,t}^{DG} = 0$.

\underline{{\bf DSO Modeling: }} One of the key responsibilities of a DSO after disruptions is to restore services as soon as possible. Given different characteristics of the loads (e.g., hospital, emergency responses) and limited resources available, the DSO may need to prioritize the restoring procedures. We assume that the DSO intends to maximize the importance of loads being served within the system while minimizing the cost of energy purchased, which can be formulated in model $(\ref{mod:DSO})$.
        \begin{subequations}
        \small{
        \label{mod:DSO}
        \begin{align}
         &\underset{\substack{\boldsymbol{p^d,v} \geq \boldsymbol{0},\\ \boldsymbol{p^s}, \boldsymbol{pf}, \boldsymbol{qf}}}{\max} && \sum_{i \in \mathcal{I}^{L}} \sum_{t \in \mathcal{T}} \omega_{i,t} p_{i,t}^d - \sum_{i \in \mathcal{I}^{\mathrm{DG}}\cup \mathcal{I}^{\mathrm{CS}}} \sum_{t \in \mathcal{T}}  \rho_{i,t} p_{i,t}^s  \label{obj:DSO} \\
		&\text{\quad \ s.t.} && \sum_{l \in \mathcal{L}} pf_{l,t} \cdot \mathrm{LT}_{l,i} - \sum_{l \in \mathcal{L}} pf_{l,t} \cdot \mathrm{LF}_{l,i} = p_{i,t}^d - p_{i,t}^s,  \ \forall i \in \mathcal{I}, t \in \mathcal{T} \label{cons:DSO_P_flow}\\
		& &&\sum_{l \in \mathcal{L}} qf_{l,t} \cdot \mathrm{LT}_{l,i} - \sum_{l \in \mathcal{L}} qf_{l,t} \cdot \mathrm{LF}_{l,i} = q_{i,t}^d - q_{i,t}^s,\  \forall i \in \mathcal{I}, t \in \mathcal{T}\label{cons:DSO_Q_flow}\\
		& && 0 \leq p_{i,t}^d \leq \bar{P}_{i,t}^d,  \ \forall i \in \mathcal{I}, t \in \mathcal{T} \label{cons:DSO_d_bound}\\
		& && q_{i,t}^d = (\bar{Q}_{i,t}^d/\bar{P}_{i,t}^d) \cdot p_{i,t}^d , \ \forall i \in \mathcal{I}, t \in \mathcal{T} \label{cons:DSO_Q_Pfactor} \\
		& &&  pf_{l,t}^2 + qf_{l,t}^2 \leq \lambda_{l,t} \cdot (S_{l}^{\mathrm{max}})^2,\ \forall l \in \mathcal{L}, t \in \mathcal{T} \label{cons:DSO_line_stat}\\
		& && v_{\mathrm{FN}_{l},t}-v_{\mathrm{TN}_l,t} \leq (1-\lambda_{l,t}) \cdot K + 2 \cdot(r_{l} \cdot pf_{l,t} + x_{l} \cdot qf_{l,t}), \ \forall l \in \mathcal{L}, t \in \mathcal{T} \label{cons:DSO_bigK1} \\
		& && v_{\mathrm{FN}_{l},t}-v_{\mathrm{TN}_l,t} \geq (\lambda_{l,t}-1) \cdot K + 2 \cdot(r_{l} \cdot pf_{l,t} + x_{l} \cdot qf_{l,t}), \ \forall l \in \mathcal{L}, t \in \mathcal{T} \label{cons:DSO_bigK2} \\
		& && (V_i^{\mathrm{min}})^2 \leq v_{i,t} \leq (V_i^\mathrm{max})^2, \ \forall i \in \mathcal{I}, t \in \mathcal{T} \label{cons:DSO_voltage_bounds}
        \end{align}}
        \end{subequations}

Objective (\ref{obj:DSO}) maximizes the weighted sum of load demand $ \sum_{i \in \mathcal{I}^l} \sum_{t \in \mathcal{T}} \omega_{i, t} p_{i,t}^d$, where the weights $\omega_{i, t}$ determine the priority of the loads $p_{i,t}^d$; $\sum_{i \in \mathcal{I}^{s}} \sum_{t \in \mathcal{T}}  \rho_{i,t} p_{i,t}^s$ is the cost of energy $p_{i,t}^s$ purchased from either DG/EV owners, who want to participate in the restoration services. Operational constraints are formulated in (\ref{cons:DSO_P_flow})-(\ref{cons:DSO_voltage_bounds}), which are adapted based on the Dist-Flow equations proposed in \cite{baran1989network, guo2019impacts}. Active and reactive power flow balances are modeled in (\ref{cons:DSO_P_flow}) and (\ref{cons:DSO_Q_flow}), respectively. $pf_{l,i}$/$qf_{l,t}$ are active/reactive line flow of distribution line $l$ ($\in \mathcal{L}$) at time $t$ ($\in \mathcal{T}$), and $q_{i,t}$ is the reactive load demand at node $i$ at time step $t$. $\mathrm{LF}_{l,i}/\mathrm{LT}_{l,i}$ are mapping matrices where $\mathrm{LF}_{l,i}/\mathrm{LT}_{l,i}$ equals to 1 if line $l$ starts/connects from/to node $i$. Constraint (\ref{cons:DSO_d_bound}) limits the served load demand $p_{i,t}^d$ to be less than the expected load demand $\bar{P}_{i,t}^d$. Constraint (\ref{cons:DSO_Q_Pfactor}) maintains the same power factor for restored loads as the ratio between expected active ($\bar{P}^d_{i,t}$) and reactive ($\bar{Q}^d_{i,t}$) load demands. Constraint (\ref{cons:DSO_line_stat}) models line capacity $S_l^\mathrm{max}$ considering binary line status $\lambda_{l,t}$, with 0 indicating line outage. Constraints (\ref{cons:DSO_bigK1}) and (\ref{cons:DSO_bigK2}) calculate the squared value of voltage $v_{i,t}$ at each node where $\mathrm{FN}_l$ and $\mathrm{TN_l}$ are connection incidence matrices representing the start and end nodes of line $l$, respectively, and $r_l/x_l$ are resistance/reactance of line $l$. $K$ is a big number to ensure that when line $l$ is disconnected at $t$ (i.e., $\lambda_{l,t} = 0$), these two constraints are always satisfied. Constraint (\ref{cons:DSO_voltage_bounds}) defines the acceptable voltage range [$V_i^{\mathrm{min}}, V_i^{\mathrm{max}}$].

\underline{{\bf CSA Modeling: }} Communicating and controlling EVs/charging stations individually is challenging for DSO. Therefore, we assume a private CSA is responsible to coordinate the power transactions between all charging stations and DS. CSA aims to maximize its profits\cite{duan2020bidding} while maintaining required charging demand of each EV. The decision making of the CSA is formulated in model (\ref{model:agg}). 

Objective (\ref{obj:Aggr}) maximizes CSA's profits, calculated as revenue of selling electricity to the DSO ($\sum_{i \in \mathcal{I}^{\mathrm{CS}}}\sum_{t \in \mathcal{T}} \rho_{i,t}p_{i,t}^{\mathrm{CS}}$) subtracting incentives  ($\sum_{r \in \mathcal{R}} \sum_{s \in \mathcal{S}} \sum_{e \in \mathcal{E}} \alpha_{rs}^e q_{rs}^{\prime e}$) and battery degradation compensation (${\sum_{i \in \mathcal{I}^{CS}} \sum_{e \in \mathcal{E}} \sum_{t \in \mathcal{T}} C_{i,r,t}^{\mathrm{deg},e}(p_{i,r,t}^e)}$) paid to EV drivers. Charging station energy provision $p_{i,t}^\mathrm{CS}$ can have negative values representing that EVs are charging instead of discharging. In this case, $ \rho_{i,t}p_{i,t}^{\mathrm{CS}}$ would be the cost for CSA to charge EVs. Power prices $\rho_{i,t}$ and incentives $\alpha_{rs}^e$ provided to EV drivers are endogenously determined by market in the proposed modeling framework. Incentives $\alpha_{rs}^e$ are based on EV type $e$, which depends on their arriving state of charge (SOC) and charging needs. For example, an EV with higher arriving SOC ($\mathrm{soc}_{r,e,T^{\text{arr}}_{r,e}}$) and lower charging needs should receive higher incentives due to their higher value to the DS restoration. We adapt the Ah-throughput counting model \cite{peterson2010lithium} to model the degradation cost of EVs ($C_{i,r,t}^{\mathrm{deg},e}(p_{i,r,t}^e)$) as a set of linear constraints. An identical strategy is adopted in \cite{guo2019impacts}, which one can refer to for details. Constraints (\ref{con:Aggr_soc_calculation})-(\ref{con:Aggr_P_CS}) specify SOC requirements that the CSA needs to fulfill. We discretize EVs into different homogeneous groups $e$ based on their travel and charging characteristics, including arriving/departing time ($T_{r,e}^{\text{arr}}$/$T_{r,e}^{\text{dep}}$), arriving SOCs ($\mathrm{SOC}_{r,e}^{\text{arr}}$), and minimum departing SOC ($\mathrm{SOC}_{e}^{\text{dep}}$). Constraint (\ref{con:Aggr_soc_calculation}) models the dynamics of the aggregated SOC ($\mathrm{soc}_{r,e,t}$) of the EVs from $r$. Battery capacity $\mathrm{Cap}^e$ is considered to normalize the charged/discharged energy of EVs $\sum_{i \in \mathcal{I}^{\mathrm{CS}}} p_{i,r,t}^e$. Constraint (\ref{con:Aggr_soc}) limits the maximum and minimum SOC ($\overline{\mathrm{SOC}}_{e}$/$\underline{\mathrm{SOC}}_{e}$) for each groups of EVs based on battery specifications, and $q_{ri(s)}^{\prime e}$ is the flow of group $e$ EVs departing from node $r$ that selected charging station $i$ (connected to node $s$ in transportation system). The arrival SOC and minimum departure SOC of EVs are constrained in (\ref{con:Aggr_soc_arrival}) and (\ref{con:Aggr_soc_dep}), respectively. Constraint (\ref{con:Aggr_P_CS}) determines the total power supply/demand ($p_{i,t}^\mathrm{CS}$) of charging station $i \in \mathcal{I}^{\mathrm{CS}}$ at time $t$ by summing the normalized charging/discharging of EVs at each station where $S^{\mathrm{base}}$ is the nominal capacity of DS. 
\begin{subequations}\label{model:agg}
\small{
    \begin{align}
        &\underset{\substack{\boldsymbol{p^{\mathrm{CS}},q^\prime,soc}}}{\max} && \sum_{i \in \mathcal{I}^{\mathrm{CS}}} \sum_{t \in \mathcal{T}} \rho_{i,t} p_{i,t}^{\mathrm{CS}} - \sum_{r \in \mathcal{R}} \sum_{s \in \mathcal{S}} \sum_{e \in \mathcal{E}} \alpha_{rs}^e q_{ri(s)}^{\prime e} - {\sum_{i \in \mathcal{I}^{CS}} \sum_{r \in \mathcal{R}} \sum_{t \in \mathcal{T}}\sum_{e \in \mathcal{E}} C_{i,r,t}^{\mathrm{deg},e}(p_{i,r,t}^e) } \label{obj:Aggr}\\
        &\text{ \ s.t.} && \mathrm{soc}_{r,e,t} = \mathrm{soc}_{r,e,t-1} - \sum_{i \in \mathcal{I}^{\mathrm{CS}}}p_{i,r,t}^e/\mathrm{Cap}^e,  \ \forall r \in \mathcal{R}, e \in \mathcal{E}, t \in (T_{r,e}^{\text{arr}}, T_{r,e}^{\text{dep}}] \label{con:Aggr_soc_calculation}
    \\
        & && \sum_{i \in \mathcal{I}^{\mathrm{CS}}}q_{ri(s)}^{\prime e}\underbar{$\mathrm{SOC}$}_{e} \leq \mathrm{soc}_{r,e,t} \leq \sum_{i \in \mathcal{I}^{\mathrm{CS}}}q_{ri(s)}^{\prime e} \overline{\mathrm{SOC}}_{e},  \ \forall r \in \mathcal{R}, e \in \mathcal{E}, t \in [T_{r,e}^{\text{arr}}, T_{r,e}^{\text{dep}}] \label{con:Aggr_soc}
    \\
      &  &&\mathrm{soc}_{r,e,T^{\text{arr}}_{r,e}} = \sum_{i \in \mathcal{I}^{\mathrm{CS}}}q_{ri(s)}^{\prime e} \mathrm{SOC}_{r,e}^{\text{arr}}, \ \forall r \in \mathcal{R}, e \in \mathcal{E}  \label{con:Aggr_soc_arrival}\\
      &  &&\mathrm{soc}_{r,e,T^{\text{dep}}_{r,e}} \geq \sum_{i \in \mathcal{N}^{\mathrm{CS}}}q_{ri(s)}^{\prime e} \mathrm{SOC}_{e}^{\text{dep}},  \ \forall r \in \mathcal{R}, e \in \mathcal{E} \label{con:Aggr_soc_dep}
    \\
         & &&p_{i,t}^{\mathrm{CS}} = \sum_{r \in \mathcal{R}} \sum_{e \in \mathcal{E}}  p_{i,r,t}^{e}/S^{\mathrm{base}}, \ \forall i \in \mathcal{I}^{\mathrm{CS}}, t \in [T_{r,e}^{\text{arr}}, T_{r,e}^{\text{dep}}] \label{con:Aggr_P_CS}
    \end{align}}
\end{subequations}

\underline{{\bf EV Drivers Modeling: }} Travel and charging behavior should be considered in modeling EV drivers. We have modeled the charging behavior of EVs based on their SOC requirements as a part of CSA responsibilities in (\ref{con:Aggr_soc_calculation}-\ref{con:Aggr_soc_dep}). In this section, we will model the routing and charging location choices of EVs in transportation system.

The utility function $U_{rs}^e$ of a driver in type $e$ selecting station $s$ is formulated in (\ref{eq:util}) \cite{guo2019impacts, TTE2021}. EV drivers make charging location choices based on four aspects: locational attractiveness $\beta_{0,s}$, travel time $-\beta_1 tt_{rs}$, charging cost/revenue from charging/discharging their stored energy $\beta_2 \alpha_{rs}^e$, and a random term $\epsilon$. In this study, we adopt a multinomial logit model for charging location choices, in which $\epsilon$ follows an extreme value distribution.
    \begin{equation}
    \small{
	    U_{rs}^e = \beta_{0,s} -\beta_1 tt_{rs} + \beta_2 \alpha_{rs}^e + \epsilon \label{eq:util}}
	\end{equation}

The destination choice of EVs ($q_{rs}^e$) and path travel time ($tt_{rs}$) are coupled, see utility function (\ref{eq:util}). To capture these couplings, we adapt the classic combined distribution and assignment (CDA) model \cite{sheffi1985urban} to model their destination choices and route choices, as shown in (\ref{mod:cda}). Notice that $\forall \tau \in \mathcal{T}^\mathrm{arr}$, we will have an CDA model to solve for the traffic patterns at time $\tau$.

The objective function in (\ref{obj:cda_obj}) consists of two parts: the first part is the summation of the area under all the link travel cost functions $tt_a(\cdot)$ (e.g., Bureau of Public Roads (BPR) function); the second part is the entropy of traffic distribution $q_{rs}^e(\ln q_{rs}^e - 1)$ and utility terms (excluding time) in (\ref{eq:util}). Objective (\ref{obj:cda_obj}) is constructed in this form to guarantee the optimal solutions of (\ref{mod:cda}) are consistent with the first Wardrop principal \cite{wardrop1952some} and the multinomial logit destination choice assumption. For technical details to prove this claim, one can refer to \cite{sheffi1985urban}. Constraint (\ref{cons:cda_v_x}) calculates link flows $\boldsymbol{v}$ by summing link flows of EVs $x_{rs}^\tau$ and conventional vehicles $\bar{x}_{rs}^\tau$  traveling at time $\tau$. Constraints (\ref{cons:cda_x_q}-\ref{cons:cda_x_bar_q}) are the vehicle flow conservation at each node for the travel demand of  EV (${q}_{rs}^e$) and conventional vehicle ($\bar{q}_{rs}^\tau$), respectively. $A$ is the node-link  incidence  matrix  of  transportation network, and $E_{rs}$ is the origin-destination (OD) incidence vector. Constraint (\ref{cons:cda_q_d}) guarantees the summation of EV traffic flow distribution to each $s$ equals to the total EV travel demand from $r$, $Q_r^e$.  The equilibrium travel time for each OD pair $rs$ can be calculated as $tt_{rs} \doteq \eta^{\tau}_{rs,r} - \eta^{\tau}_{rs,s}$, where $\eta^{\tau}_{rs,n}$ is the dual variable for  (\ref{cons:cda_x_q}).
\begin{subequations}\label{mod:cda}
\small{
	\begin{align}
		& \min_{\boldsymbol{x,\bar{x}}, 
	\boldsymbol{q} \geq \boldsymbol{0} }
		& & & &\sum_{a \in \mathcal{A}} \int_{0}^{v_a^{\tau}} tt_a(u) \mathrm{d}u + \frac{1}{\beta_1} \sum_{r \in \mathcal{R}, s \in \mathcal{S}}\sum_{e \in \mathcal{E}^{\tau}} q_{rs}^{e}\left(\ln q_{rs}^{e} - 1 - \beta_2  \alpha_{rs}^e - \beta_{0,s}\right) \label{obj:cda_obj}\\
		& \text{\quad \ s.t.} 
		& & & & \boldsymbol{v}^\tau = \sum_{r \in \mathcal{R}, s \in \mathcal{S}} \boldsymbol{x}_{rs}^\tau + \sum_{r \in \bar{\mathcal{R}}, s \in \bar{\mathcal{S}}}\boldsymbol{\bar{x}}_{rs}^{\tau}, \ \forall \tau \in \mathcal{T}^\mathrm{arr} \label{cons:cda_v_x}\\
		& & &  \hspace{-1cm}(\boldsymbol{\eta}_{rs}^{\tau})& & A\boldsymbol{x}_{rs}^\tau = \sum_{e \in \mathcal{E}^{\tau}}q_{rs}^{e} E_{rs}, \ \forall r \in \mathcal{R}, s \in \mathcal{S}, \tau \in \mathcal{T}^\mathrm{arr} \label{cons:cda_x_q}\\
        & & & & & A\boldsymbol{\bar{x}}_{rs}^{\tau} = \bar{q}_{rs}^{\tau}E_{rs}, \; \forall r \in \bar{\mathcal{R}}, s \in \bar{\mathcal{S}}, \tau \in \mathcal{T}^\mathrm{arr}\label{cons:cda_x_bar_q}\\
		& & & & &  \sum_{s \in \mathcal{S}} q_{rs}^{e} = Q_r^e, \forall r \in \mathcal{R}, e \in \mathcal{E}^{\tau}\label{cons:cda_q_d}
	\end{align}}
\end{subequations}

\underline{{\bf Market Clearing Conditions: }}  {The hourly market clearing conditions } can be stated as (\ref{eq:equi}). In a stable market, the power purchased and supplied by DSO needs to be balanced with locational power generation and power load, respectively. (\ref{eq:equi_gene}) guarantees that the total energy purchased by DSO is equal to the total energy generated at each node. { (\ref{eq:equi_char}) enforces the balance between EV flow of type $e$ demanded and supplied at each location $s$ from $r$.} Locational prices of electricity $\rho_{i,t}$ and EV incentives $\alpha_{rs}^e$ can be calculated from the dual variables for the market clearing conditions, respectively.
	\begin{subequations}
	\small{
		\begin{align}
		& (\rho_{i,t}) && p_{i,t}^s = p_{i,t}^{DG} + P_{i,t}^{\mathrm{CS}}, \; \forall i \in \mathcal{I}^{\mathrm{DG}} \cup \mathcal{I}^{\mathrm{CS}}, \ \forall t \in \mathcal{T} \label{eq:equi_gene}\\
		& (\alpha_{rs}^e) && q_{rs}^{\prime e}= q_{rs}^{e}, \ \forall r \in \mathcal{R}, s \in \mathcal{S},e \in \mathcal{E} \label{eq:equi_char}
		\end{align}
		\label{eq:equi}}
	\end{subequations}
\vspace{-0.7cm}
\section{Convex Reformulation}

The decision making of each stakeholder and market clearing conditions presented in Sections \ref{sec:MAth_modeling} are interdependent and need to be solved simultaneously to achieve the equilibrium solutions. However, due to the non-convex nature of the N-MOPEC, directly solving it is challenging. In this section, we propose an exact convex reformulation that can recover the optimal primal and dual variables efficiently. We observe that models (\ref{mod:sp})$\sim$(\ref{model:agg}) and (\ref{mod:cda}) are convex optimization problems with constraints completely separable. The objective functions of these models are almost separable except the multiplication terms of primal and dual variables in market clearing conditions (\ref{eq:equi}). This type of problem can be reformulated by linearly combining all the objective functions and constraints and applying the reverse procedures of Lagrangian relaxation on the market clearing conditions (\ref{eq:equi}) \cite{TTE2021}. Accordingly, the N-MOPEC can be reformulated as a single convex optimization problem (\ref{cons:combined}).
{\small
\begin{align} 
 \underset{\substack{(\boldsymbol{p^d,p^{DG},v,x,\bar{x},q,q^\prime})\geq \boldsymbol{0},\\
 \boldsymbol{p^s,pf,qf,p,p^{\mathrm{CS}},\mathrm{soc}}}}{ \max} & \sum_{t \in \mathcal{T}}\big(\sum_{i \in \mathcal{I}} \omega_{i,t} p_{i,t}^d - \sum_{i \in \mathcal{I}^{DG}}  C(p_{i,t}^{DG}) - \sum_{i \in \mathcal{I}^\mathrm{CS}, r \in \mathcal{R}, e \in \mathcal{E}} C_{i,r,t}^{deg,e}\big)
- { \frac{\beta_1}{\beta_2} \sum _{a \in \mathcal{A}, \tau \in \mathcal{T}^{\mathrm{arr}}} \int _{0}^{v_a^\tau} tt_a(u) \mathrm{d}u}  
- \frac{1}{\beta _2}\sum_{\tau \in \mathcal{T}^{\mathrm{arr}}}\sum_{r \in R, s \in S} \sum_{e \in \mathcal{E}^\tau} q_{rs}^e\left(\ln q_{rs}^e - 1 - \beta _0^s\right)\nonumber
  \\
  \text{s.t} & \quad (\ref{cons:DG_max}), (\ref{cons:DSO_P_flow})\sim(\ref{cons:DSO_voltage_bounds}), (\ref{con:Aggr_soc_calculation})\sim(\ref{con:Aggr_P_CS}), (\ref{cons:cda_v_x})\sim(\ref{cons:cda_q_d}),  (\ref{eq:equi}) \label{cons:combined}
  \end{align}}%
\vspace{-0.5cm}
\section{Numerical Simulation}\label{sec:Numerical}
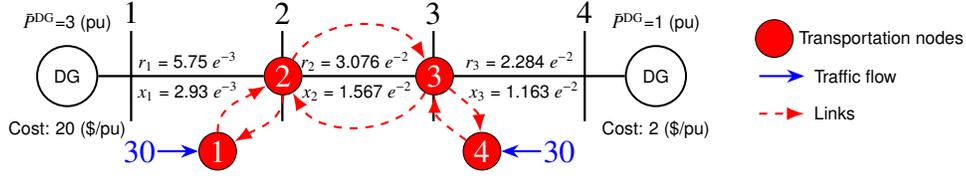
\begin{figure}
\centering
\begin{tikzpicture}[
    font=\sf \scriptsize,
    >=LaTeX,
    operator/.style={circle,draw,inner sep=-0.5pt,minimum height =0.5cm, fill=red!100, font = \large}, 
    ct/.style={circle,draw,line width = .75pt,minimum width=0.8cm,inner sep=1pt,fill=white},
    dot/.style = {circle,fill, inner sep=0.01mm, fill=black!15, node contents={}},
    arr/.style = {red, thick, dashed, ->},
    arrow/.style = {-Stealth,blue, thick},
    ]
\node[ct, name = dg1] {DG};
\node[ct, right = 7cm of dg1](dg2){DG};
\node[operator]at(9.4,0.5)(ln){} -- node[pos = 0.98, right]{Transportation nodes}(9.8,0.5);
\draw [arrow] (9.2,0) -- node[pos = 1,  right]{\textcolor{black}{ Traffic flow}}(9.8,0);
\draw [arr] (9.2,-0.5) -- node[pos = 1,  right]{\textcolor{black}{ Links}}(9.8,-0.5);
\node[operator, right = 2.2cm of dg1](l2){\textcolor{white}{2}};
\node[operator, right = 1.5cm of l2](l3){\textcolor{white}{3}};
\node[operator]at(2,-1)(l1){\textcolor{white}{1}};
\node[operator,  right= 3cm of l1](l4){\textcolor{white}{4}};

\draw [arrow] (1.2,-1) -- node[pos = 0.19, left, font=\large]{30}(l1);
\draw [arrow] (6.3,-1) -- node[pos = 0.19, right , font=\large]{30}(l4);
\draw [line width=0.30mm] (0.85,-0.6) -- node[pos = 0.98, above, font=\large]{1}(0.85,0.6);
\draw [line width=0.30mm] (2.86,-0.6) -- (2.86,-.25);
\draw [line width=0.30mm] (2.86,0.25) -- node[pos = 0.89, above, font=\large]{2}(2.86,0.6);
\draw [line width=0.30mm] (4.87,-0.6) -- (4.87,-0.25);
\draw [line width=0.30mm] (4.87,0.25) -- node[pos = 0.89, above, font=\large]{3}(4.87,0.6);
\draw [line width=0.30mm] (6.88,-0.6) -- node[pos = 0.98, above, font=\large]{4}(6.88,0.6);
\draw[line width=0.30mm] (dg1) -- node[pos = 0.405]{}(l2);
\draw[line width=0.30mm] (l2) -- node[pos = 0.405]{}(l3);
\draw[line width=0.30mm] (l3) -- node[pos = 0.405]{}(dg2);
\draw [out=60, in=120, red, thick, dashed,->]  (l2) to (l3);
\draw [out=240, in=-60, red, thick, dashed,->]  (l3) to (l2);
\draw [out=90, in=210, red, thick, dashed,->]  (l1) to (l2);
\draw [out=-90, in=30, red,thick, dashed,->]  (l2) to (l1);
\draw [out=-40, in=95,red, thick, dashed,->]  (l3) to (l4);
\draw [out=145, in=-90, red, thick, dashed,->]  (l4) to (l3);

\node[above=0.05 cm of dg1]{$\bar{P}^\mathrm{DG}$=3 (pu)};
\node[below=0.05 cm of dg1]{Cost: 20 (\$/pu)};
\node[above=0.05 cm of dg2]{$\bar{P}^\mathrm{DG}$=1 (pu)};
\node[below=0.05 cm of dg2]{Cost: 2 (\$/pu)};
\node at (1.6,0.2) (r1) {$r_1$ = 5.75 $e^{-3}$};
\node at(1.6,-0.2)(x1){$x_1$ = 2.93 $e^{-3}$};
\node[right=0.6cm of r1](r2){$r_2$ = 3.076 $e^{-2}$};
\node[right=0.6cm of x1](x2){$x_2$ = 1.567 $e^{-2}$};
\node[right=0.5cm of r2](r3){$r_3$ = 2.284 $e^{-2}$};
\node[right=0.5cm of x2](x3){$x_3$ = 1.163 $e^{-2}$};
\end{tikzpicture}
\captionsetup{labelfont=bf}
\captionsetup{justification=raggedright,singlelinecheck=false}
\caption{Four node test system}
\label{fig:four_node_system}
\end{figure}

We test the proposed models and reformulation techniques with a four-node test system shown in \textbf{Figure  \ref{fig:four_node_system}}. Nodes 2 and 3 are the connecting points of the distribution and transportation systems representing the location of charging stations. Nodes 2 $\sim$ 4 are load nodes where loads at node 3 have higher priority ($\omega_3 = 60$) compared to the other nodes ($\omega_{1} = \omega_{2} = 50$). All the distribution lines have the same capacity of $S_{l}^{\mathrm{max}} = 2$ (pu), and transportation links have the capacity of 20 vehicles/hour. EVs are in one of the three groups based on their arrival SOCs: low ($\mathrm{SOC}^{\mathrm{arr}}_{r,1}$ = 0.3), medium ($\mathrm{SOC}^{\mathrm{arr}}_{r,2} = 0.6$), and high ($\mathrm{SOC}^{\mathrm{arr}}_{r,3} = 0.8$). We assume that two traffic flows of 30 EVs/hour, with 10 EVs/hour from each group $e$, depart from transportation nodes 1 and 4 and arrive at random hours $\tau$ ($\in \mathcal{T}^\mathrm{arr}$) to charging stations. We consider losing line (1--2) unexpectedly at $t = 10$, disconnecting the important DG at node 1 from the system. The line comes back on at $t = 20$. During this period, generation from DG at node 4 is not enough to serve load demand of all the nodes, and we will considered two levels of $\mathrm{SOC}^{\mathrm{dep}}$ = 0.7 and 0.5 to investigate EVs participation in DS restoration.

\begin{figure}
\begin{minipage}[b]{0.32\linewidth}
  \centering
  \subfloat[\label{fig:case2_load_soc70}]{\includegraphics[width=\linewidth ]{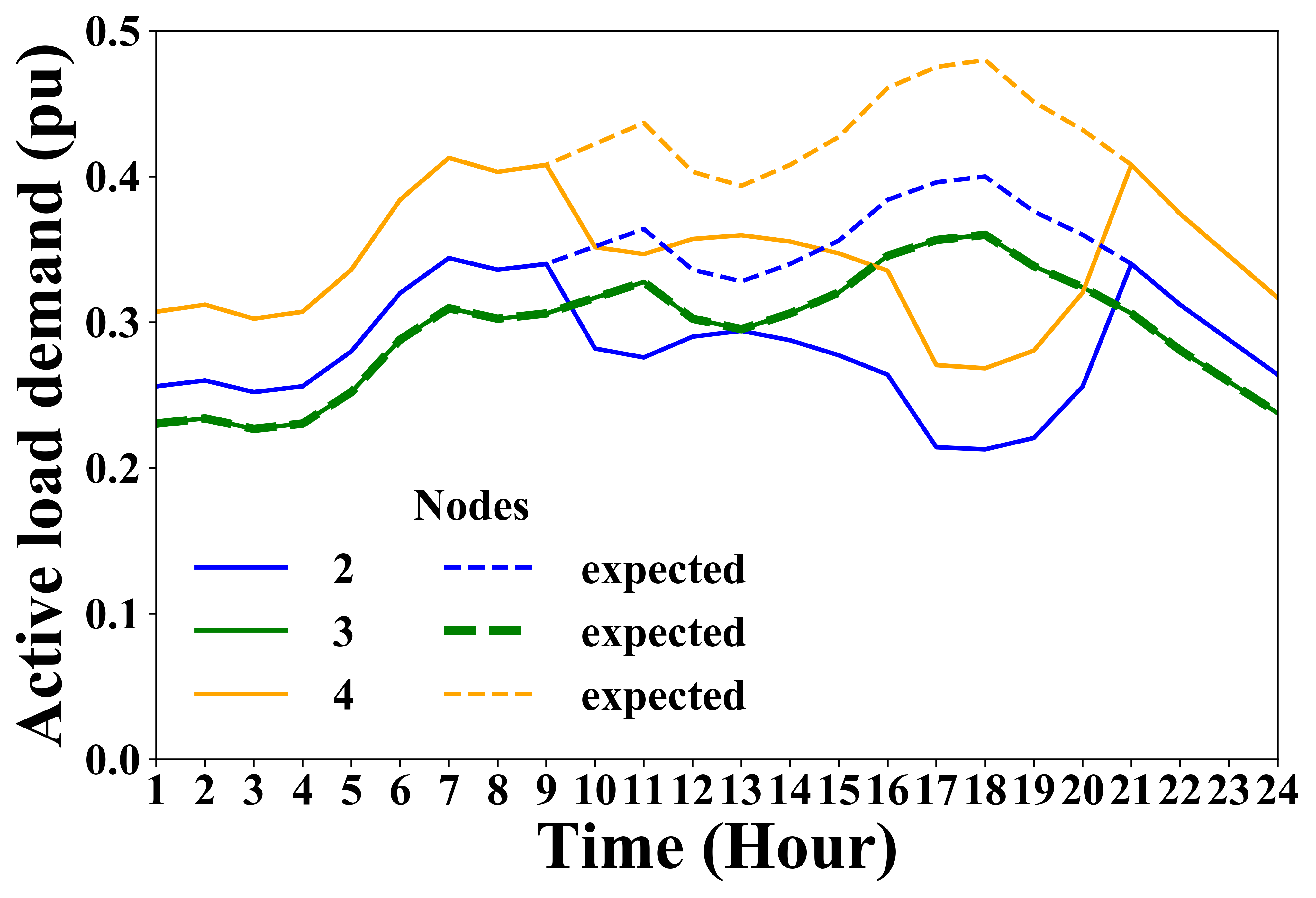}}
\end{minipage}
\begin{minipage}[b]{0.32\linewidth}
  \centering
  \subfloat[]{\includegraphics[width=\linewidth, ]{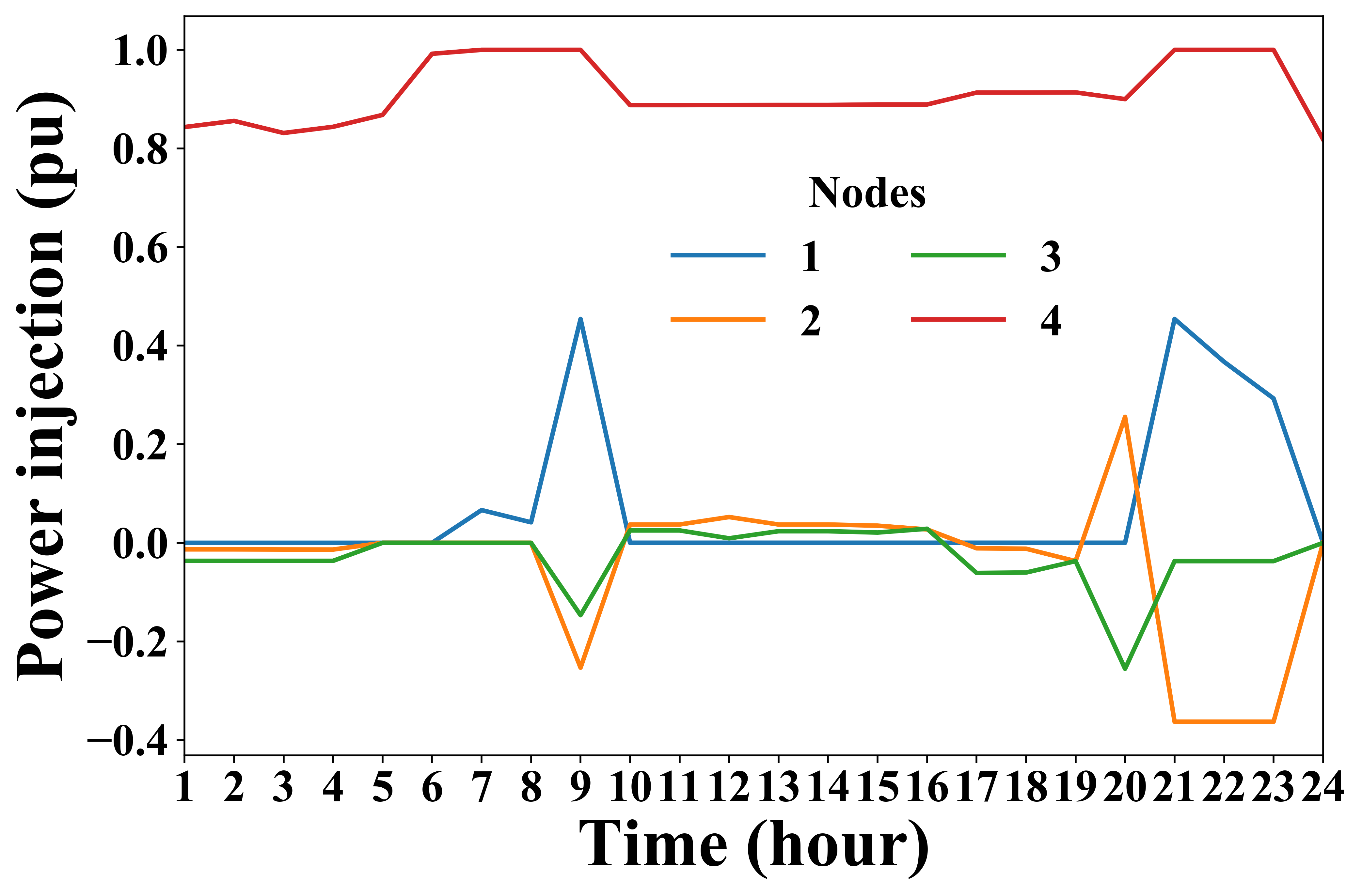}\label{fig:case2_node_power_injection_soc70}}
\end{minipage}
\begin{minipage}[b]{0.32\linewidth}
  \centering
  \subfloat[]{\includegraphics[width=\linewidth, ]{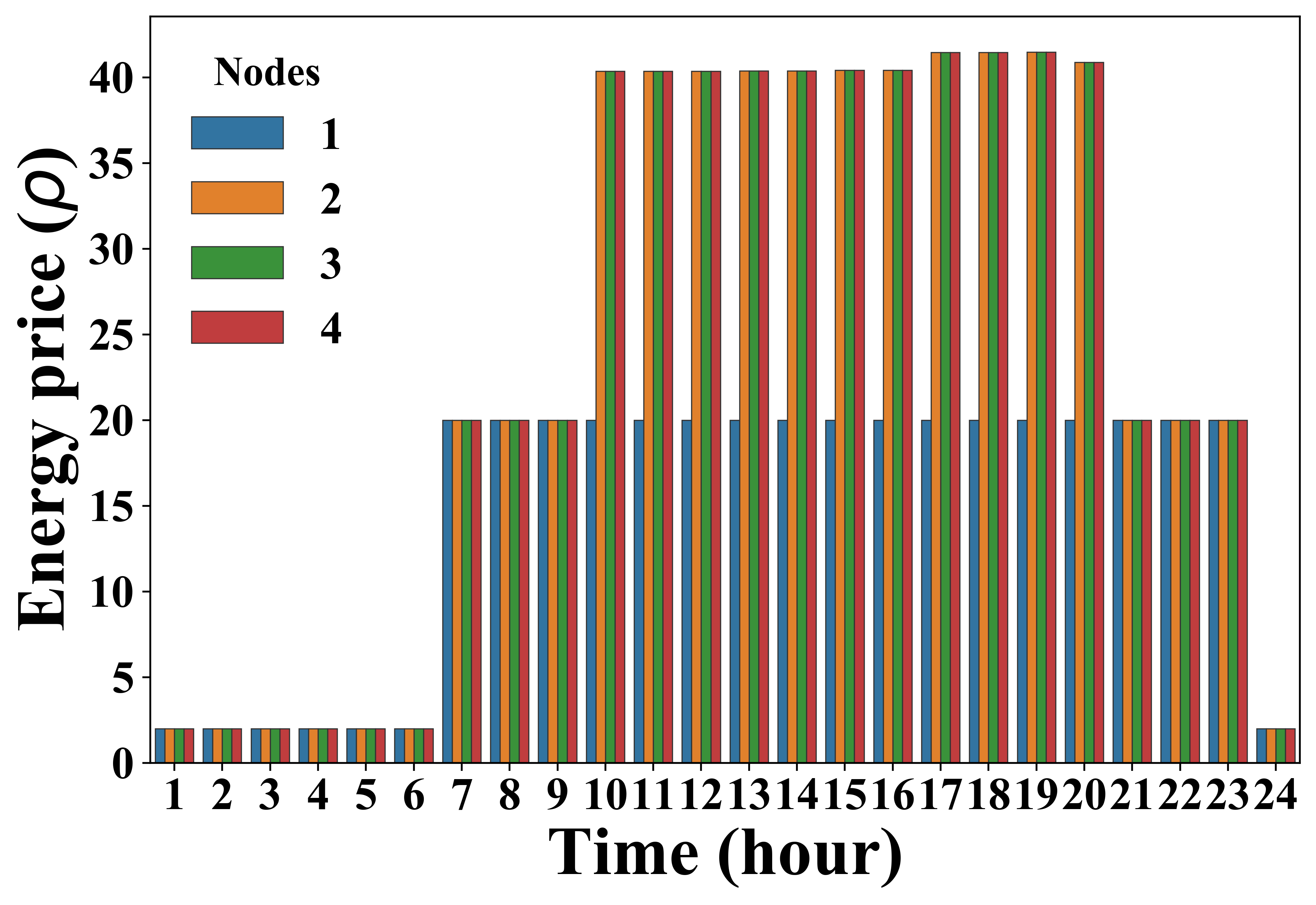}\label{fig:case2_energy_price_soc70}}
\end{minipage}

\begin{minipage}[b]{0.32\linewidth}
  \centering
  \subfloat[\label{fig:case2_load_soc50}]{\includegraphics[width=\linewidth ]{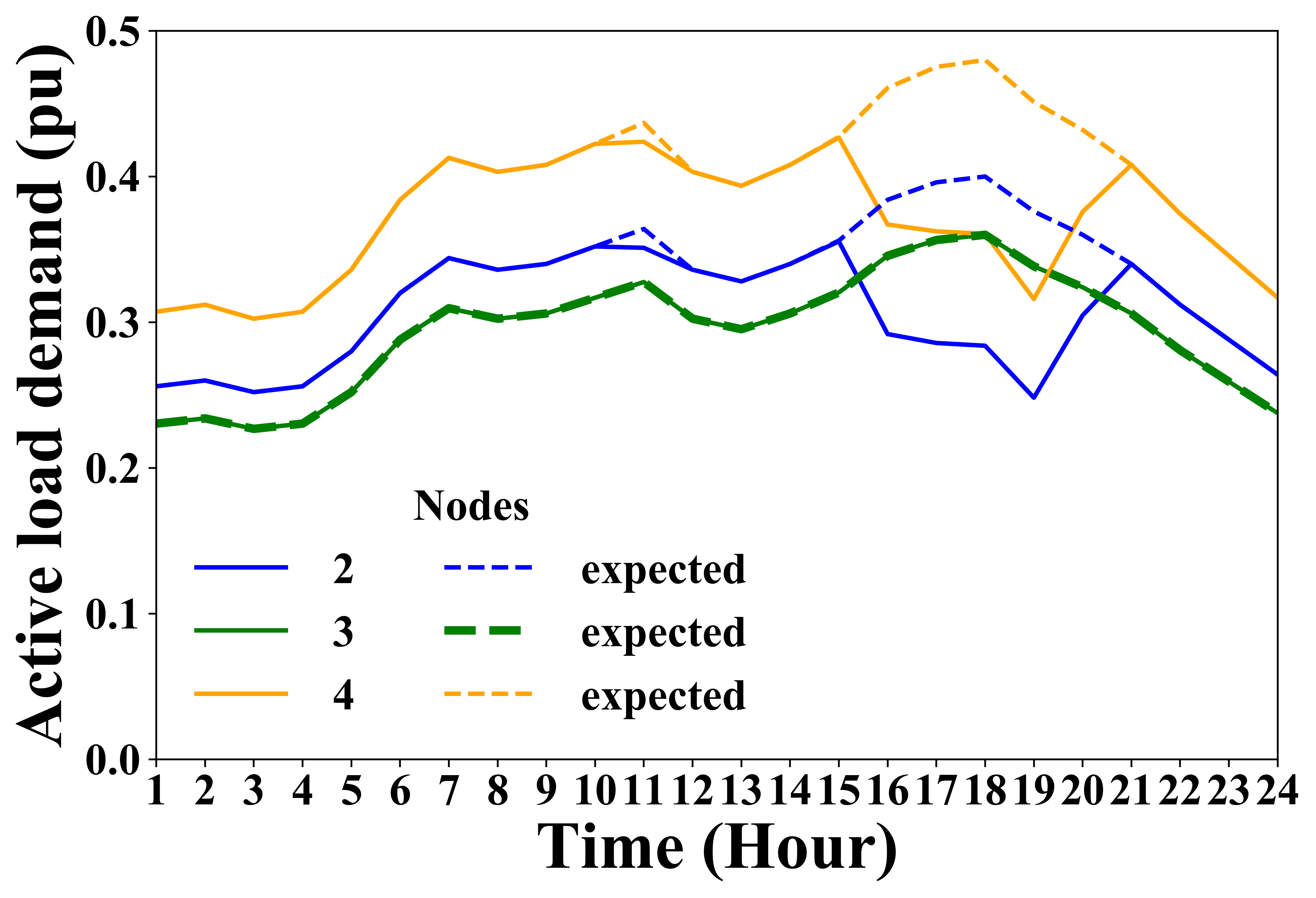}}
\end{minipage}
\begin{minipage}[b]{0.32\linewidth}
\centering
  \subfloat[]{\includegraphics[width=\linewidth ]{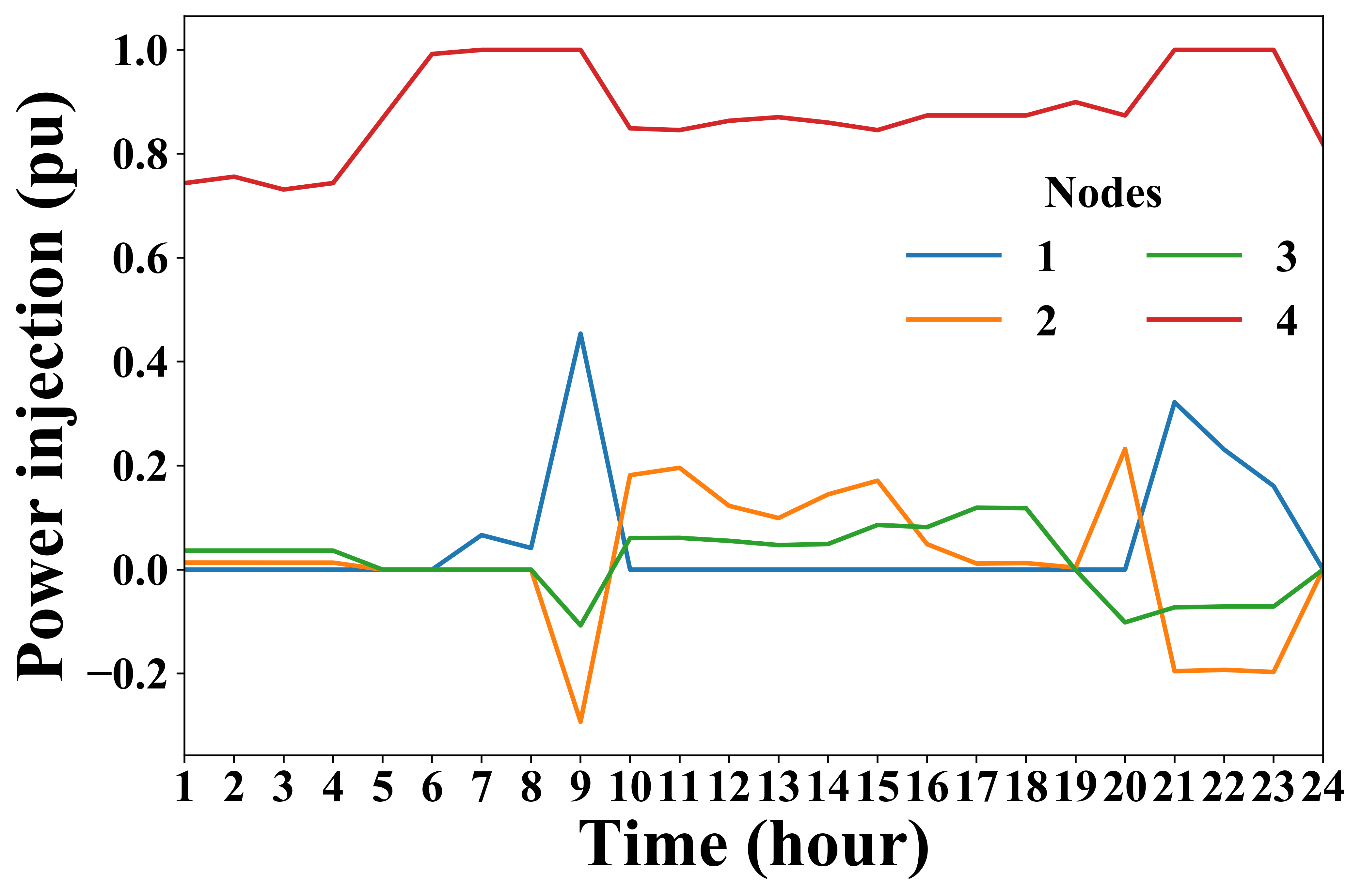}\label{fig:case2_node_power_injection_soc60}}
\end{minipage}
\begin{minipage}[b]{0.32\linewidth}
  \centering
  \subfloat[]{\includegraphics[width=\linewidth, ]{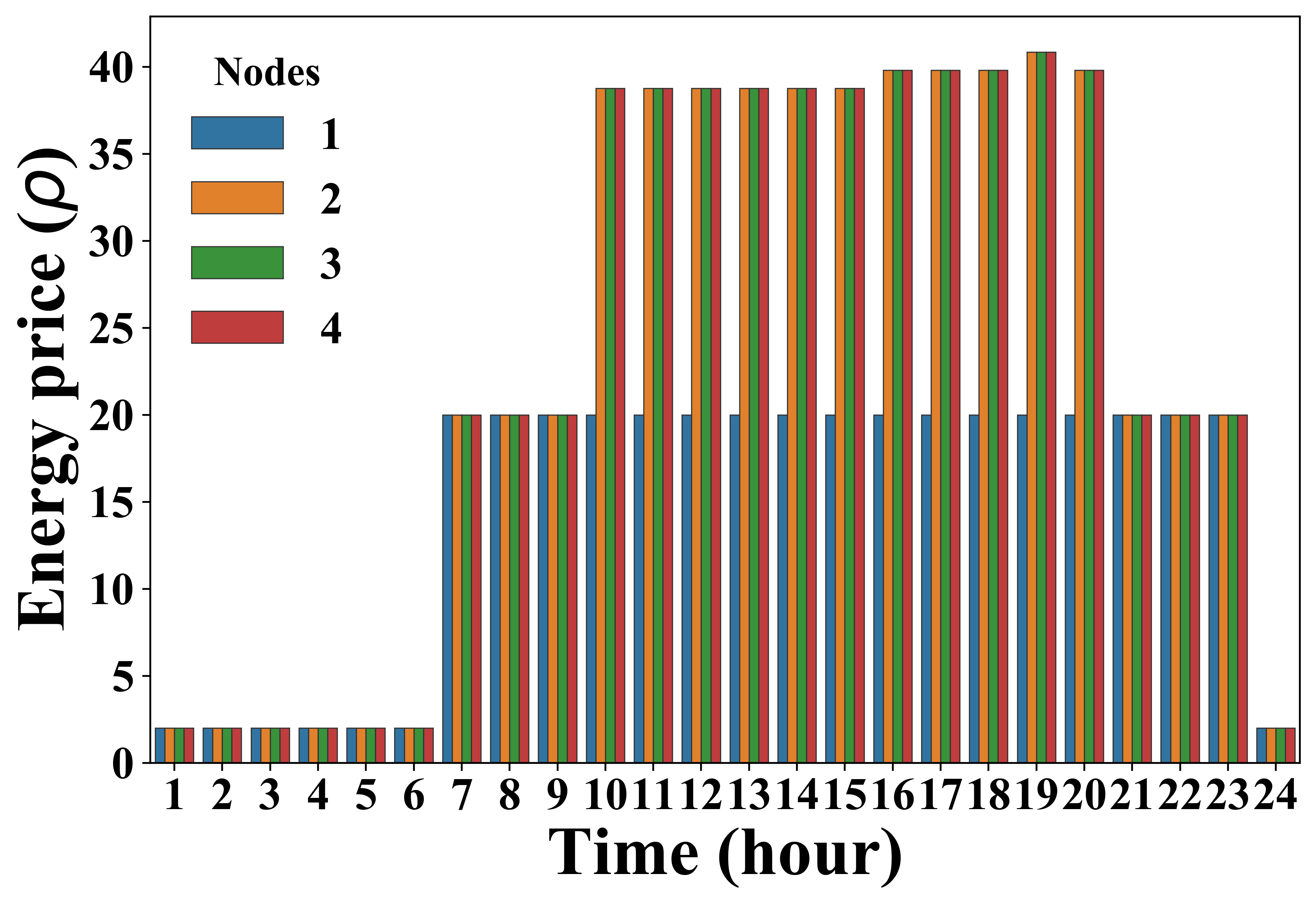}\label{fig:case2_energy_price_soc50}}
\end{minipage}
      \captionsetup{justification=raggedright,singlelinecheck=false}

  \caption{Distribution system results: Expected and picked up load (a) $\mathrm{SOC}^{\mathrm{dep}}$ = 0.7 (d) $\mathrm{SOC}^{\mathrm{dep}}$ = 0.5. Nodal power injection (b) $\mathrm{SOC}^{\mathrm{dep}}$ = 0.7 (e) $\mathrm{SOC}^{\mathrm{dep}}$ = 0.5. Nodal energy price (c) $\mathrm{SOC}^{\mathrm{dep}}$ = 0.7 (f) $\mathrm{SOC}^{\mathrm{dep}}$ = 0.5}
  \label{fig:case2_DS_results}
\end{figure}

\begin{figure}
\begin{minipage}[b]{0.24\linewidth}
  \centering
  \subfloat[]{\includegraphics[width=\linewidth ]{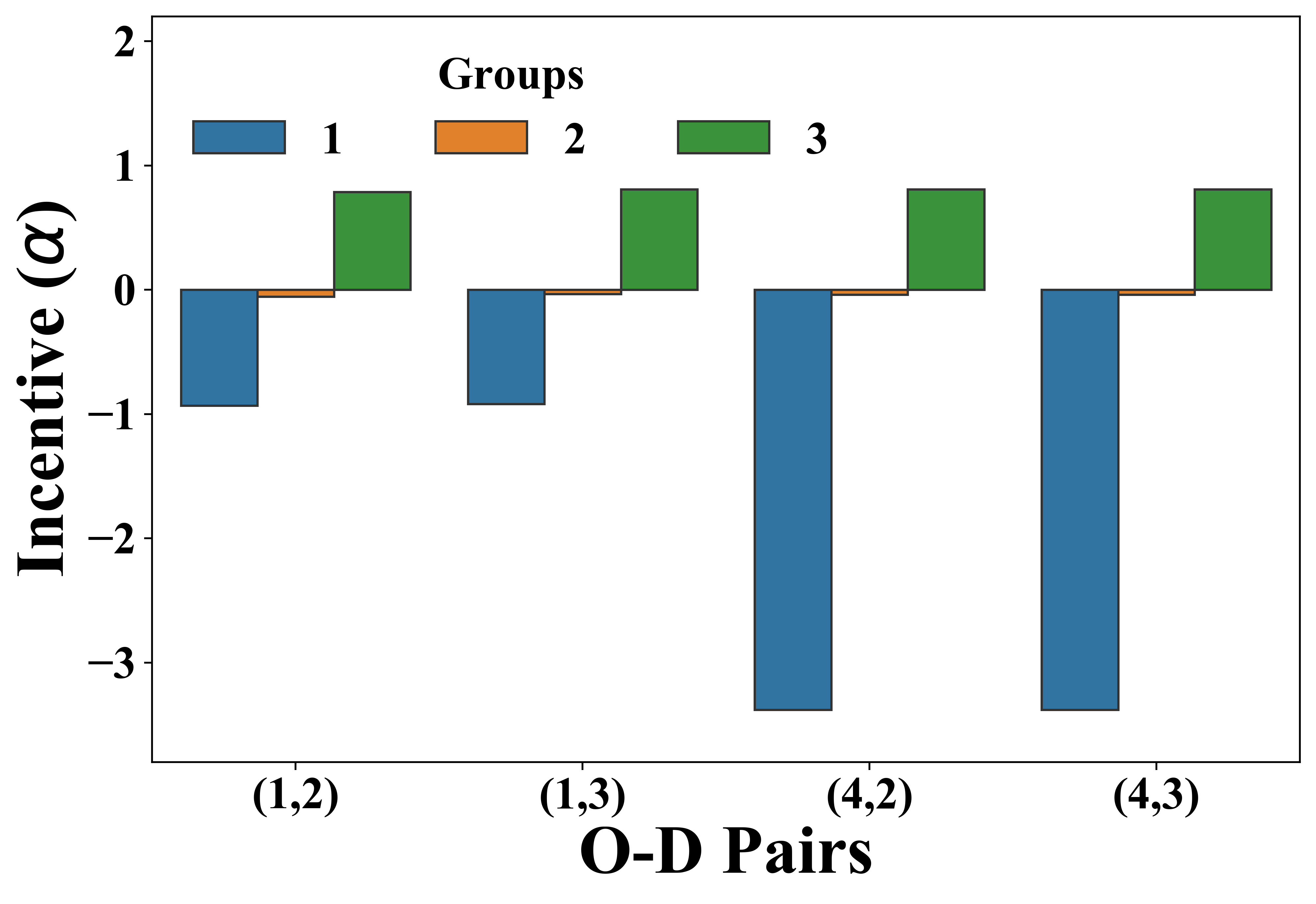}\label{fig:case2_incentive_soc70}}
\end{minipage}
\begin{minipage}[b]{0.24\linewidth}
  \centering
  \subfloat[]{\includegraphics[width=\linewidth ]{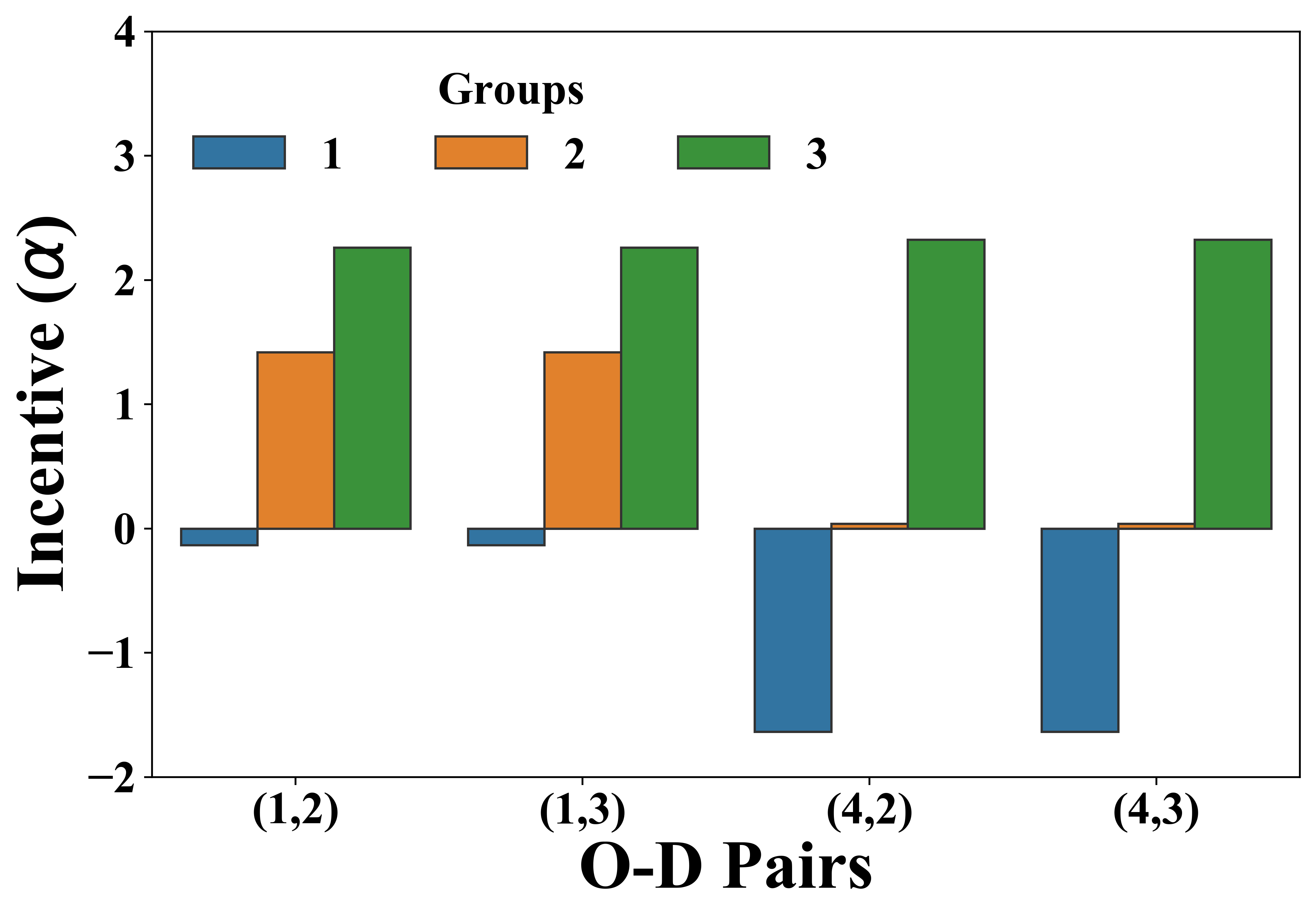}\label{fig:case2_incentive_soc50}}
\end{minipage}
\begin{minipage}[b]{0.24\linewidth}
  \centering
  \subfloat[]{\includegraphics[width=\linewidth ]{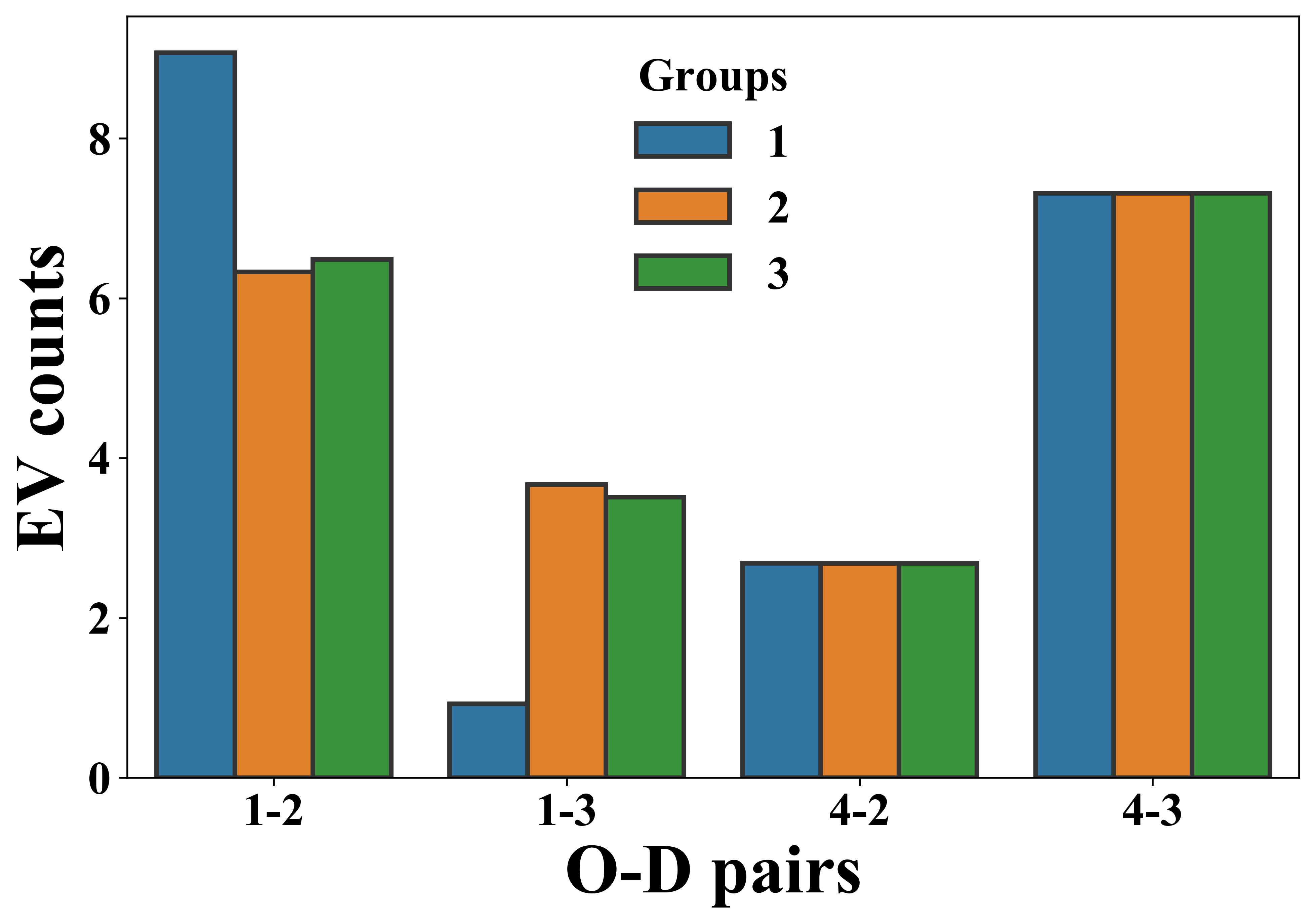}\label{fig:case2_EV_traffic_soc70}}
\end{minipage}
\begin{minipage}[b]{0.24\linewidth}
\centering
  \subfloat[]{\includegraphics[width=\linewidth ]{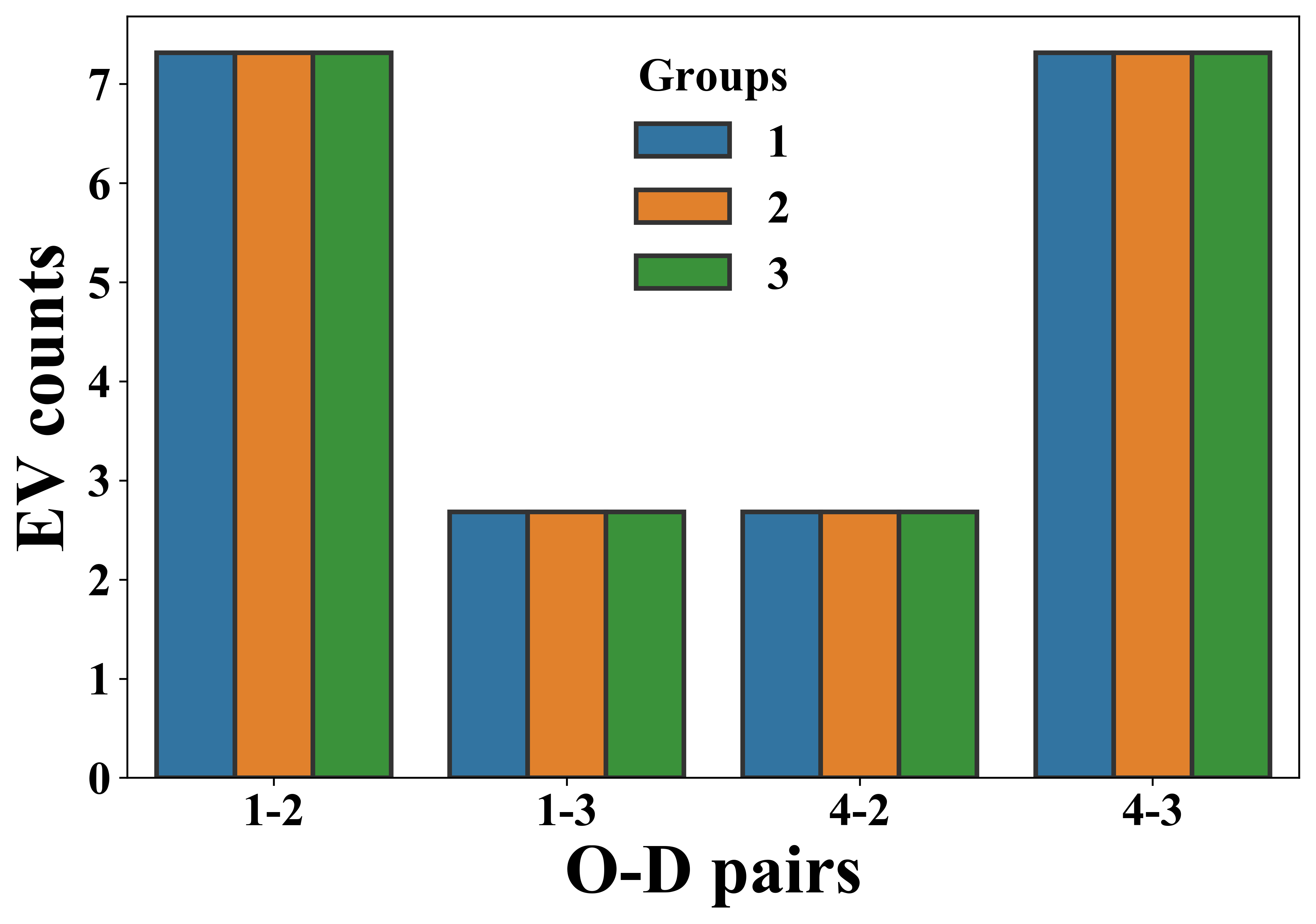}\label{fig:case2_EV_traffic_soc50}}
\end{minipage}
      \captionsetup{justification=raggedright,singlelinecheck=false}
  \caption{Transportation system results: Charging station incentives for EVs (a) $\mathrm{SOC}^{\mathrm{dep}}$ = 0.7 (b) $\mathrm{SOC}^{\mathrm{dep}}$ = 0.5. EV traffic flows (c) $\mathrm{SOC}^{\mathrm{dep}}$ = 0.7 (d) $\mathrm{SOC}^{\mathrm{dep}}$ = 0.5} 
  \label{fig:case2_transportation_results}
\end{figure}

The load pickup graphs presented in \textbf{Figure \ref{fig:case2_load_soc70}, \ref{fig:case2_load_soc50}} shows that the system is able to provide the energy for the higher priority load at node 3. With lower $\mathrm{SOC}^{\mathrm{dep}}$, more lower priority loads are picked up. The total load loss decrease from 2.316 pu to 1.056 pu when $\mathrm{SOC}^{\mathrm{dep}}$ decrease from 0.7 to 0.5. The power injection of charging stations in \textbf{Figure \ref{fig:case2_node_power_injection_soc70}, \ref{fig:case2_node_power_injection_soc60}} shows more participation of EVs with $\mathrm{SOC}^{\mathrm{dep}} = 0.5$ during the disruption compared to $\mathrm{SOC}^{\mathrm{dep}} = 0.7$. Energy prices during the disruption increase drastically compared to the normal operation for all the nodes (see \textbf{Figure \ref{fig:case2_energy_price_soc70}, \ref{fig:case2_energy_price_soc50}}) because the system has to leverage more expensive energy sources during this time. By observing the disruption periods for both $\mathrm{SOC}^{\mathrm{dep}}$ levels, we have slightly lower energy prices when charging stations are able to participate in restoring the load. Comparing \textbf{Figure \ref{fig:case2_incentive_soc70}} and \textbf{Figure \ref{fig:case2_incentive_soc50}}, EVs with lower required $\mathrm{SOC}^{\mathrm{dep}}$ receive higher incentives because they are more flexible to provide energy to the system. In addition, we see higher incentives for group 1/2 departing from node 1 than node 4, because they arrive in different time with different energy prices and/or DS restoration requirements. The charging station selection of EVs only varies slightly for $\mathrm{SOC}^{\mathrm{dep}}$ levels (see \textbf{Figure \ref{fig:case2_EV_traffic_soc70}, \ref{fig:case2_EV_traffic_soc50}}), because incentives to go to different locations for each group of EVs from the same origins are very close so that travel time is dominant.

\section{Discussion}\label{sec:conclusion}

We investigate the impact of private EVs on distribution system restoration in a N-MOPEC framework, where each stakeholder, including DG owners, DSO, EV drivers, and CSA,  maximizes their own objectives in a coupled transportation and distribution system. We reformulate the multi-agent problems as an equivalent convex optimization problem, which can be efficiently solved by commercial nonlinear solvers. Numerical results on the small-scale test system show that participation of EVs helps to reduce the load loss during restoration process, and the CSA could provide different incentives to EV drivers based on their value for the distribution system restoration. 

This work can be extended in multiple directions. First, the proposed modeling framework can be used to design optimal incentives for EVs to participate in distribution system restoration services and guide the planning and expansion of transportation and power systems. Second, stochastic modeling on renewable energy sources, EV arrival and departure time, and arrival SOC can be further examined. Third, decomposition based algorithms can be developed based on the convex reformulation to facilitate computation for extreme large-scale systems.
\setstretch{0.9}
\bibliographystyle{IEEEtran}
\bibliography{bibliography.bib}

\begin{thebibliography}{10}
\providecommand{\url}[1]{#1}
\csname url@samestyle\endcsname
\providecommand{\newblock}{\relax}
\providecommand{\bibinfo}[2]{#2}
\providecommand{\BIBentrySTDinterwordspacing}{\spaceskip=0pt\relax}
\providecommand{\BIBentryALTinterwordstretchfactor}{4}
\providecommand{\BIBentryALTinterwordspacing}{\spaceskip=\fontdimen2\font plus
\BIBentryALTinterwordstretchfactor\fontdimen3\font minus
  \fontdimen4\font\relax}
\providecommand{\BIBforeignlanguage}[2]{{%
\expandafter\ifx\csname l@#1\endcsname\relax
\typeout{** WARNING: IEEEtran.bst: No hyphenation pattern has been}%
\typeout{** loaded for the language `#1'. Using the pattern for}%
\typeout{** the default language instead.}%
\else
\language=\csname l@#1\endcsname
\fi
#2}}
\providecommand{\BIBdecl}{\relax}
\BIBdecl

\bibitem{khazeiynasab2020resilience}
S.~R. Khazeiynasab and J.~Qi, ``Resilience analysis and cascading failure
  modeling of power systems under extreme temperatures,'' \emph{J. Mod. Power
  Syst. Clean Energy}.

\bibitem{lei2018routing}
S.~Lei, C.~Chen \emph{et~al.}, ``Routing and scheduling of mobile power sources
  for distribution system resilience enhancement,'' \emph{IEEE Trans. Smart
  Grid}, vol.~10, no.~5, pp. 5650--5662, Dec. 2018.

\bibitem{liu2016ev}
H.~Liu, J.~Qi \emph{et~al.}, ``Ev dispatch control for supplementary frequency
  regulation considering the expectation of ev owners,'' \emph{IEEE Trans. on
  Smart Grid}, vol.~9, no.~4, pp. 3763--3772, 2016.

\bibitem{haggi2019review}
H.~Haggi, M.~Song, W.~Sun \emph{et~al.}, ``A review of smart grid restoration
  to enhance cyber-physical system resilience,'' in \emph{2019 IEEE Innovative
  Smart Grid Technologies-Asia (ISGT Asia)}.\hskip 1em plus 0.5em minus
  0.4em\relax IEEE, 2019, pp. 4008--4013.

\bibitem{xu2019resilience}
Y.~Xu, Y.~Wang \emph{et~al.}, ``Resilience-oriented distribution system
  restoration considering mobile emergency resource dispatch in transportation
  system,'' \emph{IEEE Access}, vol.~7, pp. 73\,899--73\,912, 2019.

\bibitem{jamborsalamati2019enhancing}
P.~Jamborsalamati, M.~Hossain \emph{et~al.}, ``Enhancing power grid resilience
  through an iec61850-based ev-assisted load restoration,'' \emph{IEEE Trans.
  Ind. Informat.}, vol.~16, no.~3, pp. 1799--1810, Jun. 2019.

\bibitem{nejad2019distributed}
R.~R. Nejad and W.~Sun, ``Distributed load restoration in unbalanced active
  distribution systems,'' \emph{IEEE Trans. Smart Grid}, vol.~10, no.~5, pp.
  5759--5769, Jan. 2019.

\bibitem{momen2020using}
H.~Momen, A.~Abessi \emph{et~al.}, ``Using evs as distributed energy resources
  for critical load restoration in resilient power distribution systems,''
  \emph{IET GENER TRANSM DIS}, vol.~14, no.~18, pp. 3750--3761, Jun 2020.

\bibitem{sun2018optimal}
W.~Sun, N.~Kadel \emph{et~al.}, ``Optimal distribution system restoration using
  phevs,'' \emph{IET Smart Grid}, vol.~2, no.~1, pp. 42--49, Oct. 2018.

\bibitem{yao2018transportable}
S.~Yao, P.~Wang, and T.~Zhao, ``Transportable energy storage for more resilient
  distribution systems with multiple microgrids,'' \emph{IEEE Trans. Smart
  Grid}, vol.~10, no.~3, pp. 3331--3341, 2018.

\bibitem{yao2019resilient}
S.~Yao, J.~Gu \emph{et~al.}, ``Resilient load restoration in microgrids
  considering mobile energy storage fleets: A deep reinforcement learning
  approach,'' \emph{arXiv preprint arXiv:1911.02206}, 2019.

\bibitem{yan2018robust}
M.~Yan, N.~Zhang \emph{et~al.}, ``Robust two-stage regional-district scheduling
  of multi-carrier energy systems with a large penetration of wind power,''
  \emph{IEEE Trans Sustain Energy}, vol.~10, no.~3, pp. 1227--1239, 2018.

\bibitem{baran1989network}
M.~E. Baran and F.~F. Wu, ``Network reconfiguration in distribution systems for
  loss reduction and load balancing,'' \emph{IEEE Power Engineering Review},
  vol.~9, no.~4, pp. 101--102, 1989.

\bibitem{guo2019impacts}
Z.~Guo, Z.~Zhou \emph{et~al.}, ``Impacts of integrating topology
  reconfiguration and vehicle-to-grid technologies on distribution system
  operation,'' \emph{IEEE Trans. Sustain. Energy}, vol.~11, no.~2, pp.
  1023--1032, May 2019.

\bibitem{duan2020bidding}
X.~Duan, Z.~Hu, and Y.~Song, ``Bidding strategies in energy and reserve markets
  for an aggregator of multiple ev fast charging stations with battery
  storage,'' \emph{IEEE trans Intell Transp Syst}, 2020.

\bibitem{peterson2010lithium}
S.~B. Peterson, J.~Apt, and J.~Whitacre, ``Lithium-ion battery cell degradation
  resulting from realistic vehicle and vehicle-to-grid utilization,'' \emph{J.
  Power Sources}, vol. 195, no.~8, pp. 2385--2392, Apr. 2010.

\bibitem{TTE2021}
Z.~Guo, F.~Afifah, J.~Qi, and S.~Baghali, ``A stochastic multi-agent
  optimization framework for interdependent transportation and power system
  analyses,'' \emph{IEEE Trans. Transp}, 2021.

\bibitem{sheffi1985urban}
Y.~Sheffi, \emph{Urban transportation networks}.\hskip 1em plus 0.5em minus
  0.4em\relax Prentice-Hall, Englewood Cliffs, NJ, 1985, vol.~6.

\bibitem{wardrop1952some}
J.~Wardrop, ``Some theoretical aspects of road traffic research,'' \emph{Proc.
  Inst. Civ. Eng.}, no. Part II, pp. 325--378, 1952.

\end{thebibliography}
\end{document}